# Structure-dependent electrical properties of graphene nanoribbon devices with graphene electrodes

## Authors

Leonardo Martini* [1-2-4], Zongping Chen [3], Neeraj Mishra [4], Gabriela Borin Barin [8], Paolo Fantuzzi [1-2], Pascal Ruffieux [8], Roman Fasel [8,10], Xinliang Feng [6], Akimitsu Narita [3], Camilla Coletti [4-5], Klaus Müllen [3], Andrea Candini [1-7]**

* Corresponding author. Center for Nanotechnology Innovation @ NEST, Istituto Italiano di Tecnologia, Piazza San Silvestro 12, 56127 Pisa, Italy

>   E-mail address: leonardo.martini@iit.it;

** Corresponding author. Istituto Sintesi Organica e Fotoreattività ISOF – CNR, via Gobetti 101 40129 Bologna, Italy

>   E-mail address: andrea.candini@isof.cnr.it

1) Centro S3, Istituto Nanoscienze−CNR, via G. Campi 213/A, 41125 Modena. Italy

2) Dipartimento di Scienze Fisiche, Matematiche e Informatiche, Università di Modena e Reggio Emilia, via G. Campi 213/A, 41125/A Modena. Italy

3) Max Planck Institute for Polymer Research, Ackermannweg 10, D-55128 Mainz, Germany

4) Center for Nanotechnology Innovation @ NEST, Istituto Italiano di Tecnologia, Piazza San Silvestro 12, 56127 Pisa, Italy

5) Graphene Labs, Istituto Italiano di Tecnologia, via Morego 30, 16163 Genova, Italy

6) Center for Advancing Electronics Dresden (cfaed) & Department of Chemistry and Food Chemistry, Technische Universität Dresden, 01062 Dresden, German

7) Istituto per la Sintesi Organica e la Fotoreattività ISOF, Consiglio Nazionale delle Ricerche CNR, via P. Gobetti 101, 40129 Bologna, Italy.

8) Empa, Swiss Federal Laboratories for Materials Science and Technology, Überland Str. 129, 8600 Dübendorf, Switzerland.

10) Department of Chemistry and Biochemistry, University of Bern, 3012 Bern, Switzerland

## Abstract

Graphene nanoribbons (GNRs) are a novel and intriguing class of materials in the field of nanoelectronics, since their properties, solely defined by their width and edge type, are controllable with high precision directly from synthesis. Here we study the correlation between the GNR structure and the corresponding device electrical properties. We investigated a series of field effect devices consisting of a film of armchair GNRs with different structures (namely width and/or length) as the transistor channel, contacted with narrowly spaced graphene sheets as the source-drain electrodes. By analyzing several tens of junctions for each individual GNR type, we observe that the values of the output current display a width-dependent behavior, indicating electronic bandgaps in good agreement with the predicted theoretical values. These results provide insights into the link between the ribbon structure and the device properties, which are fundamental for the development of GNR-based electronics.

## 1. Introduction

Organic materials for electronic applications, owning the possibility to tune the device proprieties directly from the molecular design and synthesis, have impacted the fields of semiconducting and optoelectronic devices[1]–[3]. The study of the structure-property correlation is at the heart of the continuous effort to design and realize materials with improved performances[4].

Graphene Nanoribbons (GNRs), one-dimensional stripes of graphene, are emerging as a novel class of molecular conductors, whose electrical, optical and magnetic properties are fully determined by their widths and edge structures [5]–[7]. Specifically, GNRs with zigzag edges show metallic and spin-polarized states[7]–[9], while armchair GNRs (AGNRs) present a width-dependent semiconducting bandgap [10]–[12]. The possibility to tune the electrical properties, along with the predicted intrinsic high mobility[13], makes GNRs a promising material for the next generation of nanoelectronic devices[14]. Earlier attempts to realize GNRs by top-down approaches[15]–[17] lacked the atom-scale control which is needed to fully exploit their properties. Recently, the bottom-up chemical synthesis of GNRs, which yields ribbons with uniform width and atomically precise edges, has been reported for a large variety of GNRs with different structures[18], exploiting either the solution [19],[20] or the surface[21] -assisted methods. In particular, by surface-assisted synthesis, that has been demonstrated in ultrahigh vacuum (UHV)[21]–[23] as well as by chemical vapor deposition (CVD) [24],[25], it is possible to grow flat GNRs with the desired morphology by a suitable choice of the molecular precursor[26]–[31].

Along with the development of the synthetic processes, the electrical characterization of the bottom-up fabricated GNRs is also steadily developing, from the characterization of films and networks of GNRs[32],[33] to device integration of single (or few) ribbons[34][35]. In this context, we have recently reported the use of graphene as a suitable electrode material for GNR-based FET and photosensor devices[24],[36], thus realizing all-graphene monolithic circuits[37]. In our approach the as-fabricated GNR film is transferred directly onto the graphene electrodes without the need of any further fabrication step. This method is particularly promising to maintain the intrinsic properties of the GNRs and therefore to investigate the predicted dependence on their structure, an important step towards the use of GNRs in electronics devices. In this work, we investigate the electrical behavior of CVD-grown AGNRs films with different widths and hence different electronic bandgaps. Namely, we used 5-AGNR,[38] 9-AGNR,[39] and 5n-AGNR,[38] where 5 and 9 indicate the number of carbon atom along the ribbon width and the sample labelled 5n-AGNR is a mixture containing mainly 10-AGNR and also some amounts of 5-, 15-, and 20-AGNRs (see section 2.2 for details). Moreover, the 9-AGNR sample was prepared also by the UHV method[35], to compare the behavior of GNRs with the same width but different length. The AGNR film samples are contacted employing graphene-based electrodes with narrow (<50 nm) gaps. The small distance between the electrodes and the direct transfer of the as-grown GNR films without further fabrication steps allow us to limit the device variability induced by inter-ribbon junctions and polymer residues. We analyzed ~50 devices for each GNR sample and observed FET behaviours with high output currents (up to hundreds of nA with 1 V of applied bias), which also revealed that the mean electrical conductivity of the final device is inversely related to the value of the specific GNR bandgap. Furthermore, we demonstrate that the GNR length plays a role in the device behavior, as longer GNRs lead to improved output currents, as a consequence of the larger contact area and the further reduced number of ribbon-to-ribbon junctions. Our work demonstrate the correlation between the GNRs structure and the resulting device electrical properties, in accordance with the theoretical predictions, and highlights the use of graphene in GNR-based devices for electronic and optoelectronic applications.

## 2. Experimental methods

### 2.1. Fabrication of graphene electrodes and characterization of the devices

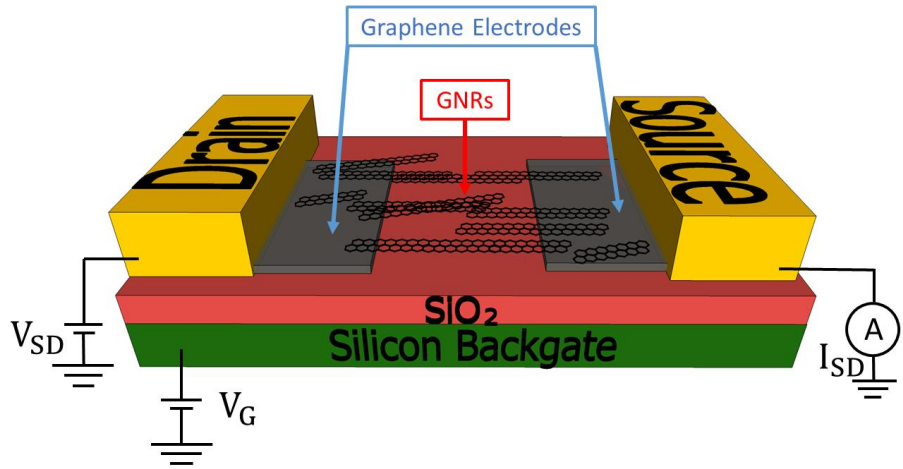

*Figure 1 Schematic view of our devices: metal (Cr/Au) electrodes contact the device made of two graphene electrodes bridged by one or more GNRs (not to scale). The device is realized on a substrate of doped Si covered by 300 nm of SiO$_2$ that is used as a global backgate.*

Figure 1 shows a schematic view of our graphene/GNR devices: two graphene electrodes (contacted with Au) are bridged by a film of AGNRs and the underlying doped Si covered by 300 nm of SiO$_2$ is used as a global backgate. For the realization of this scheme, we proceed as follows: the starting material is a large area graphene sheet grown by chemical vapour deposition (CVD) on copper[40] and then transferred on the SiO$_2$-Si substrate, following a reported procedure[40]. The realization of the electrical contacts on graphene is done with electron beam lithography (EBL) and metal evaporation (5 nm of chromium and 50 nm of gold). The patterning of graphene is realized with EBL and oxygen plasma (20 SCCM of O$_2$ at 50 W for 30 seconds) in a reactive ion etching (RIE) chamber to remove the unprotected graphene. Initially, the bare graphene devices are annealed at 375 K for at least 1 hour in vacuum. A typical transfer curve is shown in Figure 2a ($I_{SD}$ vs $V_{Gate}$).

We then open a gap in the graphene flake in order to use it as the lateral electrodes to contact the GNR film. To do so, we employed a modified version of the electroburning procedure that we have previously reported in [41]: the $V_{SD}$ bias voltage is continuously increased until the current start to drop (see Figure 2b), meaning that a rupture is starting to form in the graphene sheet. To avoid a complete breakdown of the device the voltage is restored to 0 V in less than 100 ms when the current falls below a threshold value. After this procedure, the device behaves as an "open" circuit and no signature of tunnelling current is observed up to a voltage bias $V_{SD} = 2$ V, (Figure 2c), implying that the gap width is at least ~10 nm[41]. We stress that the absence of tunnelling current is fundamental to correctly assess the electrical signal arising from the GNRs contacting the graphene electrodes and not simply from tunnelling between nanometer-spaced electrodes[42].

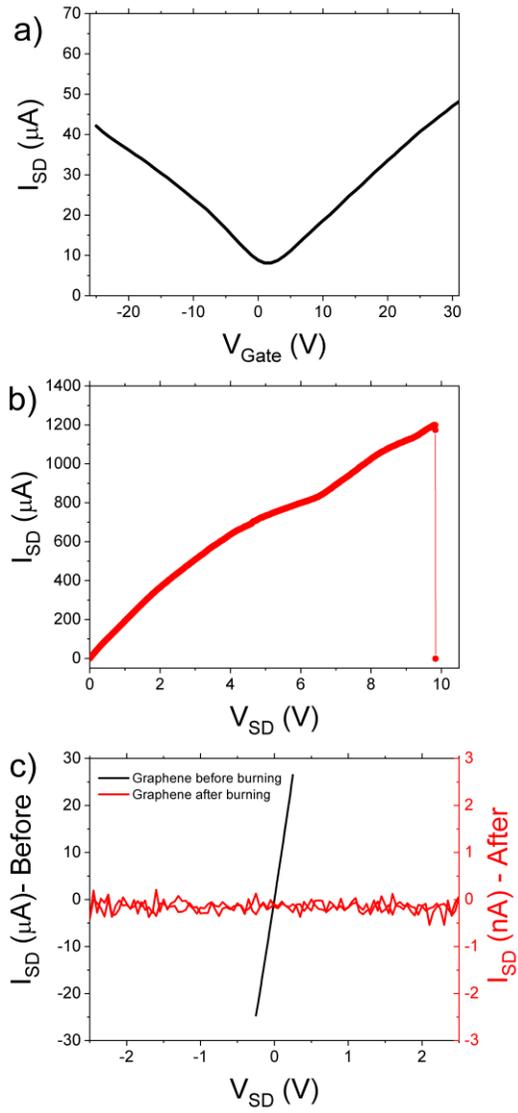

*Figure 2 a) Gate voltage dependence of the graphene source-drain current $I_{SD}$ vs $V_{gate}$ before the gap opening. b) $I_{SD}$ vs $V_{SD}$ curve during the electroburning procedure: the bias voltage is increased until the current drops and then immediately reset to zero within < 100 ms to avoid the complete rupture of the device. c) $I_{SD}$ vs $V_{SD}$ curves before (black, left Y-axis) and after (red, right Y-axis) the burning process shown in b): note the different scales for the current axis. The absence of a tunneling current signal after the electroburning procedure is a fingerprint of a device completely open (gap of at least 10 nm)*

The morphological characterization of the device after the fabrication and prior to the GNR transfer is shown in Figure 3, where the open gap between the two graphene electrodes is visible. The width of the aperture is measured to be 10-20 nm in its thinnest part that extends for ~100 nm. This procedure and all the electrical measurements are performed in a probe-station (LakeShore PS-100) using a dual channel source meter (Keithley 2636b) that provides both source-drain bias and the backgate potential. All the measurements are performed in the 2-terminal sensing mode and in vacuum ($10^{-4}$ Torr).

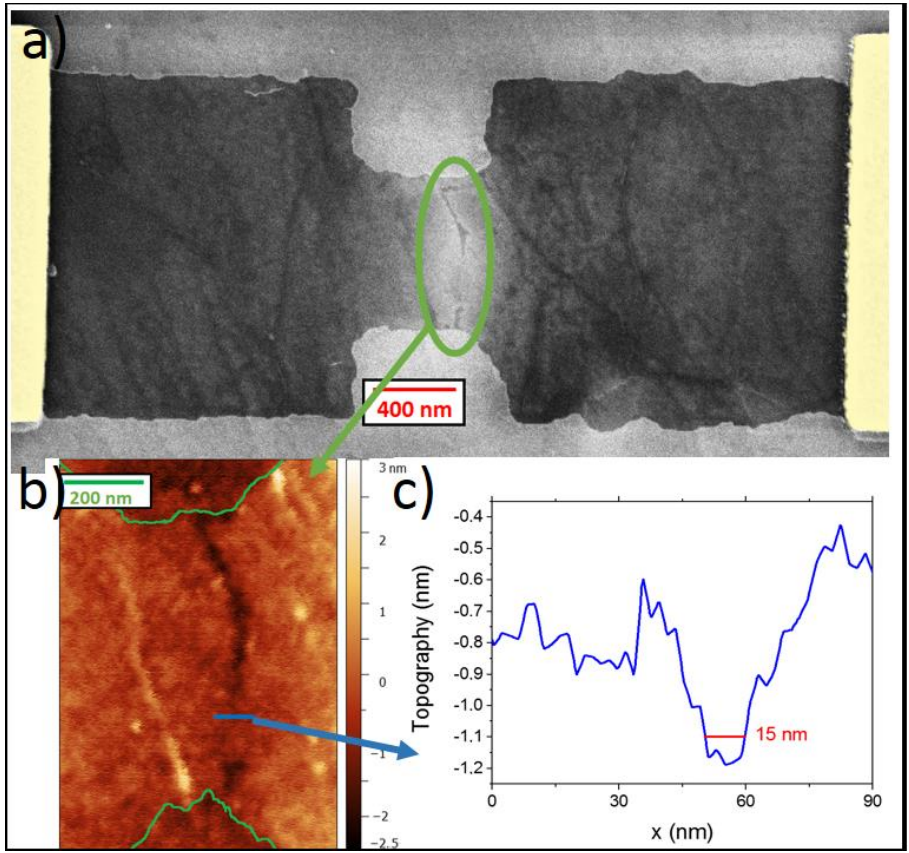

*Figure 3 a) False color SEM image of a device after the fabrication procedure: yellow: lateral metal (Cr/Au); dark grey: graphene; light grey: substrate. The open gap is visible in the middle of the flake. Its width falls below the SEM visibility in its thinnest part. b) AFM topography image (acquired with a VEECO Autoprobe CP in the tapping mode) of the part highlighted in panel a). The contour of the graphene flake is marked in green. c) Topography profile of the blue line shown in panel b).*

## 2.2. Graphene nanoribbon synthesis and transfer

Structurally defined armchair GNRs were synthesized by the on-surface CVD and UHV methods through polymerization and graphitization of dihalogenated monomers on a metal surface[24],[38],[39]. The synthesis method allows the growth of GNR with well-defined edge structure and tunable width by the suitable choice of the starting molecular monomer. The GNR samples for the first set of experiments in this study were fabricated by the CVD method into a tube furnace under gas flow of Ar (500 sccm) and $H_2$ (100 sccm) with pressure of ~1.5 mbar. 9-AGNR was prepared starting from 3',6'-dibromo-1,1':2',1''-terphenyl (**DBTP**) as the monomer (see Figure 4) at an annealing temperature of 400 °C[39] and 5-AGNR from 3,9 (10)- dibromoperylene (**DBP**) by setting the annealing temperature at 350 °C.[38] By increasing the annealing temperature, 5-AGNR can undergo lateral fusion into wider GNRs[38]. In particular, fixing the annealing temperature at 600 °C, the resulting GNRs are a mixture of 10-AGNR and non-negligible amounts of 5-AGNR, 15-AGNR and 20-AGNR (see the reaction scheme in Figure 4). The width-specific radial breathing-like mode (RBLM) of Raman spectra in Figure 5 indicates the presence of GNRs with multiple widths after the annealing of the 5-AGNR to 600 °C. It is noted that the intensity of Raman signals can vary depending on the excitation wavelength and absorption of the specific GNR at this wavelength, and thus cannot be simply used to assess their composition. More detailed characterization by means of optical spectroscopy, scanning probe (AFM, STM) and optical terahertz photoconductivity can be found in our previous reports[38],[39]. In particular, it is observed that the GNR films are made by a dense layer of GNRs with average length of about 10 nm, with the longest extending up to 35 nm[38,39].

For comparison 9-AGNRs were also synthesized by surface-assisted polymerization and cyclodehydrogenation in ultrahigh vacuum (2 x $10^{-10}$ mbar) from 3',6'-diiodine-1,1':2',1''-terphenyl (**DITP**) on Au(111)/mica surface (Phasis, Switzerland) following a reported procedure[43]. The Au(111)/mica was cleaned by repeated cycles of sputtering (1 kV Ar+ for 10 min) and annealing (at 470 °C for 10 min). In the next step the monomer was sublimed (70 °C) during 2 minutes onto the Au(111) surface kept at room temperature, resulting in approximately 0.5 monolayer coverage. After deposition, the substrate was annealed up to 200 °C (for 10 min) to induce the polymerization reaction, followed by annealing at 400 °C (for 10 min) to planarize the polymers via cyclodehydrogenation, resulting in 9-AGNRs [43]. In Figure 5b we compare the Raman spectra of both the CVD and UHV grown 9-AGNRs, highlighting the similar structural quality between the two films. For a more detailed characterization of the UHV 9-AGNR film, including scanning probes (STM) measurements, we refer to our previous report.[43, 51] We note that the STM measurements indicate that the resulting GNRs are significantly longer than the CVD-grown ribbons, with an average length of ~45 nm.

To detach the GNR film from the Au/mica substrate and transfer it onto the target substrate with the pre-defined graphene electrodes we proceed as follows (the procedures being similar for the CVD and UHV samples): the GNRs were first covered with a poly(methyl methacrylate) (PMMA) film to supply mechanical support and protect the GNRs. The mica was then detached from the substrate in a bath of hydrofluoric acid (40 wt.%) solution (hydrochloric acid 38% for the UHV sample). The gold below the GNRs was removed by a commercial gold etchant (Sigma-Aldrich) and cleaned in ultra-pure water several times. The GNR/PMMA film was then transferred on the target device with the pre-fabricated graphene electrodes and the PMMA finally removed by hot acetone washing.

After the transfer, the GNRs layer connects the two graphene electrodes, thus forming the active channel of our FET devices. The conductive paths are made by channels where the GNRs bridge directly the two electrodes, in parallel with ones where ribbon-ribbon junctions are present. The direct paths with lower resistance likely carry most of the charge current. Taking into account the size of the gap between the graphene electrodes (10 - 50 nm in its thinnest part) and the average length of the GNRs as deduced by STM on the gold substrate (~10 nm and ~45 nm for the CVD- and for the UHV-grown samples, respectively[39,39,43]), we estimate the number of junctions to be <5 for the CVD GNR and almost negligible for the longer (UHV-grown) GNRs. Importantly, the transferred GNR samples do not undergo any further fabrication processes, which is important to avoid any additional contamination and maintain the intrinsic properties of the GNR, which are related to the different structures. The graphene/GNRs devices were then re-inserted in the probe-station and annealed at 375 K for 1 h in vacuum before the electrical transport characterizations.

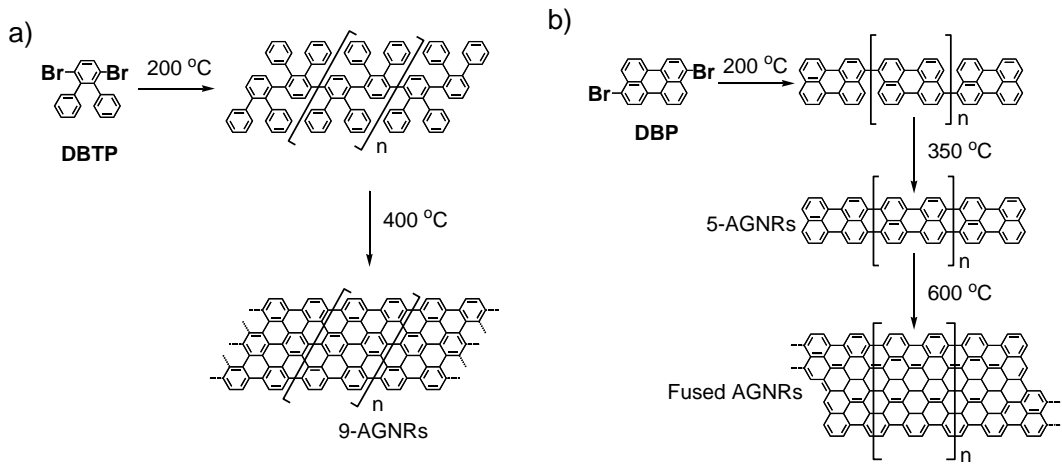

*Figure 4 Synthetic scheme toward 9-AGNRs (a) and 5-AGNRs (b) via surface-assisted dehalogenation and polymerization of monomer DBTP (a) and DBP (b), followed by cyclodehydrogenation. Further annealing of the 5-AGNRs to higher temperature of 600 °C results in the fused AGNRs with multiple widths.*

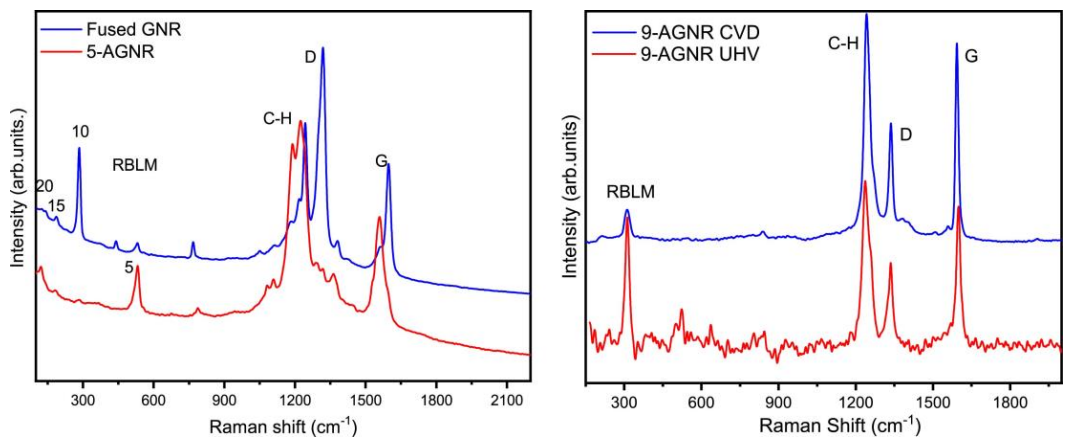

*Figure 5 a) Typical Raman spectra of 5-AGNRs grown at 350 °C and the fused wider AGNRs annealed at 600 °C. The numbers 5, 10, etc. indicate the position of the RBLM peaks of respective AGNRs, indicating the presence of GNRs with multiple widths in the fused GNR sample. b) Typical Raman spectra of 9-AGNRs grown by CVD and UHV conditions(excitation wavelength: 785 nm).*

In total, we have fabricated more than 500 graphene electrode devices that we used for the three samples of CVD-grown GNRs. After the transfer of the GNRs, we define the following criteria to assess the effective presence of the nanoribbon to bridge the gap between the graphene electrodes: the $I_{SD}$ current must be at least 500 pA with an applied source-drain voltage $V_{SD}$ of 1 V and must be gate-tunable in the range $V_{Gate} = \pm 50$ V. In total, we have found 45 working devices with 9-AGNR, 50 devices with 5-AGNR and 40 devices with 5n-AGNR with a success rate of ~20-30%. Regarding the UHV 9-AGNR, we found 38 working devices starting from 84 graphene electrode devices with a slightly higher success rate of 45%.

## 3. Results and Discussions

### 3.1. General characteristics of GNR devices contacted with graphene electrodes

Figure 6 shows a typical example of the common characteristics found in our GNR devices with graphene electrodes. For simplicity, here we show the data collected from devices with 9-AGNRs, remarking that the main features that we are discussing in this section are independent from the specific GNR type employed. We will address in the next session the differences related to the use of different GNR widths. The room-temperature source-drain current vs. source-drain bias ($I_{SD}$ vs. $V_{SD}$) curves in Figure 6a show a non-linear and asymmetric behaviour, that is a common feature in electrical devices made of bottom-up synthesized GNRs[34],[36],[44]. The reason is that the contact resistance between the GNR and the electrodes is high, with the presence of a significant Schottky-type injection barrier leading to a non-Ohmic behaviour. The output curves show a small hysteresis as it is visible from the curves in Figure 6a.

The transfer characteristic curve $I_{SD}$ vs $V_{Gate}$ at a fixed bias of $V_{SD} = 1$ V is shown in Figure 6b, where we plot the average of three consecutive sweeps. Also in this case we observe a small opening between the forward and backward curves, mostly within the electrical noise level. We observed a p-type field effect transistor (FET) behaviour, in analogy with previous reports where the same nanoribbons are contacted by metal electrodes[35]. Since in the latter the transfer of the GNRs was performed following a slightly different process, without the use of the supporting PMMA layer[35], we conclude that the observed type of doping is not directly related on the specific fabrication procedure. The on/off ratio is typically > 10 with highest values of ~1000. In order to prevent leakage from the gate, we limited the use of gate potential higher than 50 V in absolute value. However, we stress that in principle it is possible to achieve higher values for the output current simply by further decreasing the applied gate potential. On the other hand, by increasing $V_{Gate}$ to further positive values, the current reaches the lower limit of our measurement apparatus (tens of pA).

Figure 6c shows the effect of changing the temperature on the transfer characteristic curve of a similar device: from liquid nitrogen (77 K) up to 375 K, there is no significant change in the behaviour of the devices. The absence of a temperature dependence suggests that the transport mechanism at the injection barrier between the GNR and the graphene electrodes is dominated by tunnelling (Simmons type) mechanism instead of thermionic emission over the barrier[45]. Such a predominance of tunnelling effect was already reported in GNR-based devices[35] and in carbon nanotubes with non-optimized metal contacts[46].

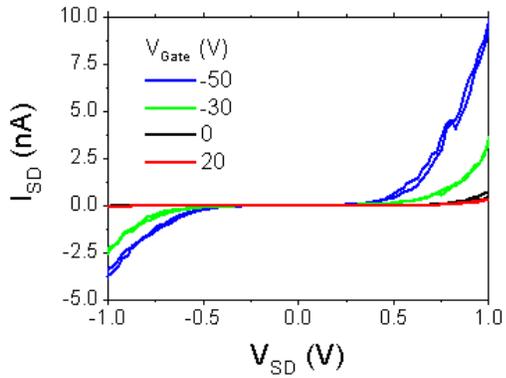
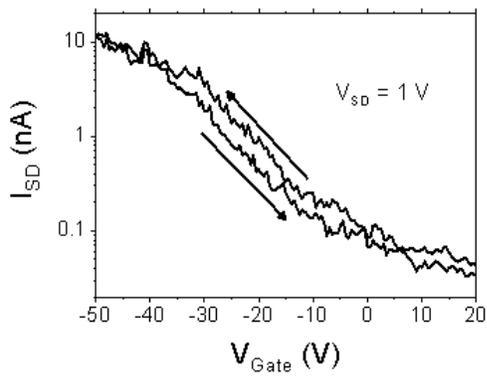
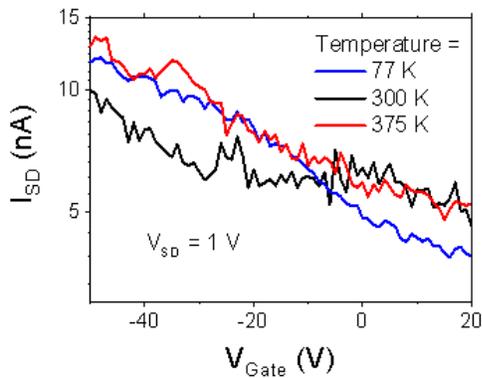

*Figure 6 a) Characteristic $I_{SD}$ vs $V_{SD}$ of our GNR devices: the non-linearity and asymmetry of the curves can be explained by the presence of a relevant barrier (Schottky-type) at the interface graphene electrode-GNRs. b) Corresponding transfer curve of a device, measured at a fixed $V_{SD}$ = 1 V: the device shows a p-type FET behaviour with an on/off ratio ~ 300. c) Transfer curves ($I_{SD}$ vs $V_{gate}$) of a similar device at different temperatures, fixing $V_{SD}$ = 1 V: the device behavior is nearly temperature independent in the investigated range, suggesting the presence of a high electrical barrier at the contact interface, which is overcome by a tunnelling mechanism (Simmons type), while the contribution from the thermionic emission mechanism is negligible.*

### 3.2. Comparison between GNR of different widths

The main goal of our work is the comparison between the behaviour of devices using different types of GNRs as the channel material. In particular, by characterizing tens of devices for each type of GNR, we have found that the mean value of the output currents depends on the specific GNR used. Figure 7a-c shows, for each type of GNR tested, the distribution of the measured $I_{SD}$ (in logarithmic scale) taken with $V_{SD} = 1$ V and $V_{Gate} = -50$ V, which is the state of maximum conductance. Here the histograms are defined as the number of tested devices with an $I_{SD}$ value within a certain range (shown in the x-axis). It is clearly visible that the current distribution significantly changes by changing the specific GNR in the device.

In order to assess in a more quantitative way the differences in the conductance properties, we fit the current distributions by assuming a Gaussian-type function, taking into account the possible device-device variations, such as differences in the effective gap length, number of GNR effectively contacted and the mechanical contact between the GNR and the graphene electrode. Because of these possible mechanisms of variations, the resulting distributions are rather broad, spanning over few orders of magnitude. This is consistent with other GNR based devices employing metal contacts[25],[35]. The corresponding fit parameters (mean value $x_0$ and standard deviation σ) are reported in Figure 7a-c and are plotted in Figure 7d. We note that the current mean value changes by almost two orders of magnitude between the 9-AGNR and the 5-AGNR. The devices made with the 5n-AGNR present a conductance slightly smaller than the 5-AGNR.

Our results can be rationalized by considering the theoretically predicted electronic bandgap values of the different GNRs[10]: since the behaviour of our devices is dominated by the contact resistance at the GNR/graphene junction, we expect that smaller-bandgap GNR will lead to a smaller injection barrier and hence higher output currents[34]. According to the theoretical calculations within the GW approximation[10], 9-AGNR is predicted to have a bandgap significantly higher than the 5-AGNR (2.1 eV vs 1.7 eV, respectively), which is in agreement with our findings. The decrease of conductivity shown by the 5n-AGNR sample can be explained by the simultaneous presence, along with the 10-AGNR, of 5- and 15- AGNRs, as a result of the high annealing temperature used during their growth as explained in Section 2.2[38]. Indeed, while the 10-AGNR (calculated bandgap ~3 eV) should be more insulating than the 9-AGNR, the contemporary presence of the lower-bandgap 15- and 20-AGNR (calculated bandgap

~1.5 eV and ~1.0 eV respectively) leads to only a moderate decrease of the conductivity with respect to the 5-AGNR devices. Our results follow a similar trend as observed in previous reports based on THz photoconductivity measurements[38,39]. While the latter experiments are performed on as-grown GNRs, probing the intrinsic mobility of the charge carriers, here we study actual electronic devices, where extrinsic effects and in particular the presence of the contacts cannot be neglected. The similarity between the observations further corroborates our simple model, since both the injection barrier and the charge carrier effective mass depend on the electronic bandgap of the GNRs[10]. Overall, these findings demonstrate the intimate link between the GNR structure and their electrical properties, also in operating devices.

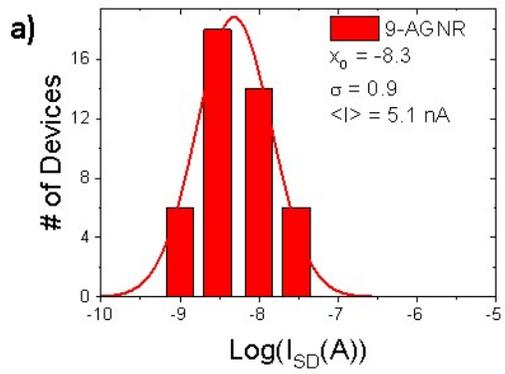
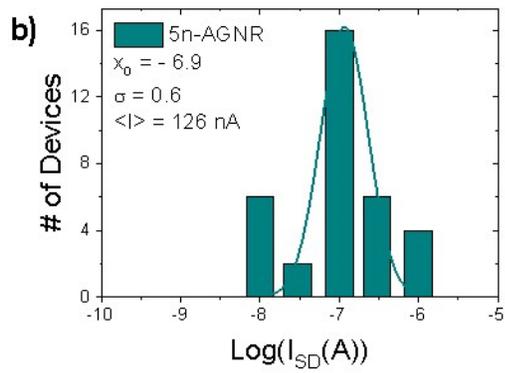
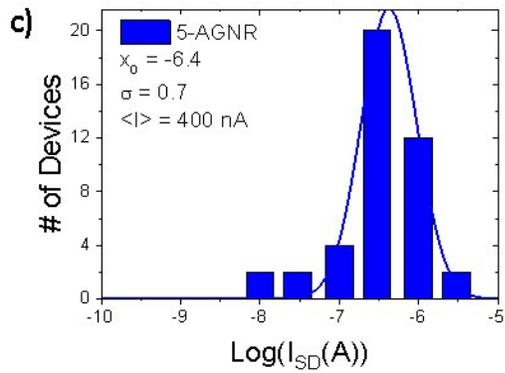
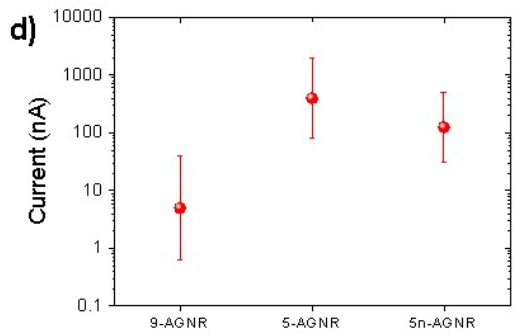

*Figure 7 a-c) Histrograms of the number of devices with a measured source-drain current value $I_{SD}$ in the corresponding interval ($V_{SD}$ = 1 V; $V_{Gate}$ = -50 V). The current values are in logarithmic scale. The distributions were fitted with a Gaussian-type function $y = \frac{A}{\sigma\sqrt{\{2\pi\}}} e^{2\left(\frac{x-x_0}{\sigma}\right)^2}$ d) Output current mean values and standard deviations for the GNRs type analyzed, as a results of the fit in a-c.*

### 3.3. Comparison between GNRs grown in UHV and CVD

Our results indicate that higher values of the output currents can be achieved by employing GNR with lower bandgap. However, the presence of non-linear I-V curves shown in Figure 6a and the relatively high resistance values found in our devices (best values of ~1MΩ with $V_{SD}$ = 1V) suggest that further engineering is needed in order to reduce the contact resistance and make the GNR/graphene devices suitable for electronic applications. The electrical contact at the interface is indeed a critical issue on devices from low-dimensional systems[47] and in particular for ultra-narrow GNRs, where the contact area is limited by the width of only few nanometers. In principle, the contact area can be increased by extending the length of the GNRs, which can be achieved by employing GNRs grown under UHV conditions. Indeed, as already discussed in Section 2.2, while the average length of 9-AGNR grown by CVD is around ~10 nm (with the longest up to 35 nm)[39] 9-AGNR grown under UHV are usually longer, with an average length that can be extended up to ~45 nm[43] when using the DITP molecular precursor as in the present work (see Section 2.2 for more details). Moreover, since the distance between our graphene electrodes is < 50 nm, most of the GNRs of the UHV-grown film will bridge directly the two contacts, thus avoiding the presence of ribbon-ribbon junctions, which is also expected to lead to a reduced device resistance.

In Figure 8 we report a comparison of the results obtained using UHV-grown 9-AGNR with those shown in Figure 7a (CVD-grown 9-AGNR). An increase of the value of the output current is clearly detected. By fitting the distributions assuming a Gaussian type dispersion as previously discussed, we found that the mean value of the currents is ~5 nA for the shorter (CVD-grown) GNRs while it is >10 nA for the longer (UHV-grown) ones. We ascribe this improvement of the electrical behaviour to the increased contact length, along with the reduced number of inter-ribbon junctions. It is worth reminding here that the two growth methods lead to GNR films with the same width and comparable structural quality as confirmed by characterization with different techniques[15],[35],[39]. Indeed, changing the GNR width has still a major role in the final device properties, as clearly pointed out by our work.

Interestingly, the output current distribution that we found using the 9-AGNR grown under UHV conditions is in line with the results obtained using Pd electrodes with a similar spacing (20 nm), but with a much thinner $SiO_2$ gate dielectric layer[35] (50 nm while here we are using 300 nm). The use of a dielectric layer as thin as possible has been shown to be of significant importance to enhance the GNR electrical behaviour by improving the current transparency through the injection barrier at the GNR/contact interface[25]. This points out the potential of using graphene electrodes for contacting other low-dimensional materials[48]–[50]. The advantages of graphene are expected to be even more effective on ultra-short devices, because of the reduced screening effect of graphene as compared to metals[47].

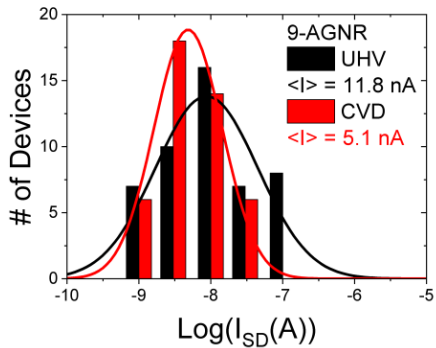

*Figure 8 Output current distributions measured for devices based on 9-AGNR grown with the UHV (black) and CVD (red) methods. Measurement conditions are $V_{SD}$ = 1 V and $V_{Gate}$ = - 50 V. Solid lines are a fit assuming a Gaussian type distribution. The corresponding mean values are ~5 nA (red) and ~12 nA (black).*

## 4. Conclusions

In this work, we investigated the correlation between structure and electrical properties in GNR-based devices, showing the possibility to tune the electronic characteristics by changing the GNR width and/or length. More specifically, we showed that the conductance of the devices depends on the GNR width and hence on the magnitude of the electronic bandgap, proving that it is possible to control the device behaviour by engineering the GNR properties directly from the synthesis. While the dependence of the GNR bandgap on their width has been investigated by various experimental techniques[38],[39], in our work we focused on a direct demonstration of a working device. We employed graphene as the contact electrodes, which allowed us to realize the final devices without any further fabrication process after the GNR transfer, thus keeping as intact as possible the intrinsic properties of the GNRs.

Furthermore, we showed an improvement of the conductance by optimizing the GNR and the electrode geometry, specifically by using longer GNR grown under UHV conditions. The observed improvement of the conductance was ascribed to the increased contact length at the graphene/GNR interface and to the reduced contribution from ribbon-ribbon junctions, leading to an enhancement of the output current. Realistic steps are therefore possible to improve the electrical behaviour of GNR-based devices, opening the possibility to use GNRs in electronic and optoelectronic functional devices.

## Acknowledgements.

We thank M. Affronte for helpful discussions and the use of the probe station experimental equipment. This work has been partially supported by the Italian Ministry for Research (MIUR) through the FIR grant RBFR13YKWX, the European Community through the FET-Proactive Project "MoQuaS"(N.610449) and Horizon 2020 research and innovation programme under GrapheneCore1 (No. 696656) and GrapheneCore2 (No. 785219) and the Max Planck Society G.B.B, P.R and R.F acknowledge the Swiss National Science Foundation under Grant No 20PC21_155644, the European Union's Horizon 2020 research and innovation programme under grant agreement number 785219 (Graphene Flagship Core 2), and the Office of Naval Research BRC Program under the grant N00014-12-1-1009 for financial support.